\documentclass[aps,prb]{revtex4}
\usepackage{amsmath,amssymb}
\usepackage{graphicx}
\usepackage{dcolumn}
\usepackage{bm}
\usepackage{color}
\usepackage{relsize}
\newcommand{\mbb}{\mathbb}
\newcommand{\mc}{\mathcal}

\newcommand{\tet}{\texttt}
\newcommand{\pr}{\partial}

\begin{document}

\title{Floquet engineering of titled and gapped Dirac materials}


\author{Andrii Iurov$^{1}$\footnote{E-mail contact: aiurov@mec.cuny.edu, theorist.physics@gmail.com},  
Liubov Zhemchuzhna$^{1,2}$,
Godfrey Gumbs$^{2,3}$, 
Danhong Huang$^{4,5}$,
Kathy Blaise$^{1}$ and Chinedu Ejiogu$^{1}$
}

\affiliation{
$^{1}$Department of Physics and Computer Science, Medgar Evers College of City University of New York, Brooklyn, NY 11225, USA\\ 
$^{2}$Department of Physics and Astronomy, Hunter College of the City University of New York, 695 Park Avenue, New York, New York 10065, USA\\ 
$^{3}$Donostia International Physics Center (DIPC), P de Manuel Lardizabal, 4, 20018 San Sebastian, Basque Country, Spain\\ 
$^{4}$US Air Force Research Laboratory, Space Vehicles Directorate, Kirtland Air Force Base, New Mexico 87117, USA\\ 
$^{5}$Center for High Technology Materials, University of New Mexico, 1313 Goddard SE, Albuquerque, New Mexico, 87106, USA\\
$^{6}$US Military Academy at West Point, 606 Thayer Road, West Point, New York 10996, USA
}

\date{\today}

\begin{abstract}

\par
\medskip 
We have established a rigorous theoretical formalism for Floquet engineering, or investigating and eventually tailoring most crucial electronic properties of tetragonal molybdenum disulfide (1T$^\prime$-MoS$_2$), by applying an external high-frequency dressing field 
in the off-resonant regime. It was recently demonstrated that monolayer semiconducting1T$^\prime$-MoS$_2$ may assume a distorted tetragonal structure which exhibits tunable and gapped spin- and valley-polarized tilted Dirac bandstructure.  From the viewpoint of electronics, 1T$^\prime$-MoS$_2$ is one of the most technologically promising nanomaterials and a novel representative of an already famous family of transition metal dichalcogenides. The obtained dressed states strongly depend on the polarization of the applied irradiation and reflect the full complexity of the initial low-energy Hamiltonian of non-irradiated material. We have calculated and analyzed the obtained electron dressed states for linear and circular types of the polarization of the applied field focusing on their symmetrical properties, anisotropy, tilting and bandgaps, as well as topological signatures. Since a circularly polarized dressing field is also known to induce a transition into a new state with broken time-reversal symmetry and a non-zero Chern number, the combination of these topologically non-trivial phases and transitions between them could reveal some truly unique and earlier unknown phenomena.
\end{abstract}

\maketitle

\section{Introduction} 
\label{sec1}
Floquet theory, or even Floquet engineering, which describes electronic behavior of a wide range of quantum-mechanical systems under a
periodic field, \cite{kibis2020floquet, kibis2018floquet} has become extremely popular over the last several years. It is important to 
note that recent experimental advances in optical and microwave physics, laser technology and emergent technical applications for condensed-matter quantum optics have enabled the experimental verification of such theoretical predictions and their applications in
actual optoelectronic devices.\,\cite{cheng2019observation,weitenberg2021tailoring,wang2018floquet,dehghani2014dissipative,nakagawa2020wannier}

\medskip
The effect of the off-resonance dressing field strongly depends on the polarization of this field. 
 Circularly polarized dressing field is known to open a bandgap and break the time-reversal symmetry in graphene.
An important technical challenge addressed by Floquet engineering is the confining ballistic electrons
within a specific spatial region of an optoelectronic device which is directly related to 
Klein paradox \,\cite{iurov2020klein} in our material and formation of the localized states 
\,\cite{pereira2006disorder,gumbs2014strongly, schnez2010imaging,castro2008localized,gumbs2014revealing} due to an induced energy bandgap.\,\cite{kibis2010metal, iurov2011anomalous}
In contrast, a linearly polarized optical field induces an in-plane anisotropy and could affect the sequential-tunneling
current of doped electrons for a non-zero polarization angle.\,\cite{ibarra2022dirac,iurov2022optically} 
A gap could also be opened in high-intensity field regime of the circularly polarized irradiation.\,\,cite{sandoval2020floquet}

\medskip
The electron-photon dressed states been studied in a variety of two-dimensional materials:\,\cite{castro2022floquet, oka2018floquet} nanotubes,\,\cite{kibis2021floquet,hsu2006floquet,sentef2015theory} graphene,\,\cite{kibis2010metal, calvo2011tuning, vogl2021floquet, dal2017one} silicene,\,\cite{iurov2017exchange} transitional metal 
dichalcogenides,\,\cite{kibis2017all} dice lattice and $\alpha-\mathcal{T}_3$,\,\cite{iurov2019peculiar,dey2019floquet, dey2018photoinduced, weekes2021generalized,tamang2021floquet} various types of nanoribbons\,\cite{wang2021floquet, tahir2016floquet, iurov2021tailoring},  anisotropic phosphorene\,\cite{iurov2017exploring} and others.\,\cite{calvo2020floquet, calvo2015floquet, iurov2013photon,mojarro2020dynamical}
The effect of circularly polarized dressing field was examined in silicene, one of the isotropic limiting
cases for 1T$^\prime$-MoS$_2$. It was found that a circularly polarized field breaks the equivalence of the two
valleys and, most surprisingly, could either increase or decrease the bandgap depending on its initial
value in a stark contrast to any known material. \cite{kibis2017all} It was also demonstrated that the off-resonant field 
could substantially modify transport properties \cite{iurov2022optically,kristinsson2016control,iurov2020quantum}, excitonic behavior \cite{iorsh2022floquet} and topological signatures of a two-dimensional lattice.\,\cite{dey2020unconventional, dey2019floquet, dal2015floquet, iorsh2017optically}

\medskip 
\medskip

A molybdenum disulfide MoS$_2$ is one of monolayer transition-metal dichalcogenides with
a distorted tetragonal structure.\,\cite{tang2017quantum,zheng2016quantum,fei2017edge}
In general, the monolayer semiconducting MoS$_2$ structure may assume a trigonal prismatic coordination of the metal atoms (2H with hexagonal symmetry), as well as octahedral coordination (1T with tetragonal symmetry) or 1T$^\prime$ with a distorted and the most exotic \,\cite{liu2018phase, mehmood2021two} which is exactly the subject of our study.

This is the one of the most innovative two-dimensional Dirac materials which is thermodynamically
stable and hence easily synthesized in its semiconducting phase. It was also theoretically predicted
to demonstrate a strong quantum spin Hall effect and a great potential for optoelectronic and other
applications.In the presence of an external transverse electric field, this material exhibits valley-spin-polarized Dirac
bands and a phase transition between the topological insulator and a regular band insulator similarly to
silicene. The energy band structure also demonstrates a special type of anisotropy which is referred to
as tilted Dirac bands.\,\cite{gomes2021tilted,tan2021anisotropic}

\medskip
1T$^\prime$-MoS$_2$ exhibits unique features, such as tunable anisotropy, bandgap, spin- and valley-polarized states and
coexistence of different topological phases similarly to silicene.\,\cite{tabert2013valley,
kara2012review, zhao2016rise, drummond2012electrically,ezawa2012valley} Some of these properties are determined by the value of
external electric field and thus could be controlled, while the others are fixed by an internal spin-orbit
coupling gap and other lattice parameters of 1T$^\prime$-MoS$_2$. The zero-gap limit of 1T$^\prime$-MoS$_2$ bands is found
in 8-Pmmn borophene which was intensively studied overt the recent years because of a strong
anisotropy in all of its crucial physical properties.\,\cite{mannix2015synthesis,
lopez2016electronic,islam2018magnetotransport,paul2019fingerprints,tan2021signatures} Very recently, a plasmonic gain was 
reported in current-biased tilted Dirac nodes \cite{park2022plasmonic} Their optical properties 
and optical conductivity have been also thoroughly investigated.\,\cite{mojarro2021optical,tan2021anisotropic} 
Other monolayer materials with similar tilted band structure, such as TaIrTe$_4$, TaCoTe$_2$ or  $\alpha-SnS_2$  have been recently
synthesized.

\medskip 
\medskip

\par
\medskip
The remaining part of the present paper is organized as follows: in Sec.~\ref{sec2} we review some crucial properties of the low-energy Hamiltonian, dispersions and the corresponding electronic states of 1T$^\prime$-MoS$_2$: tilting and anisotropy, direct and indirect bandgaps. Next section \ref{sec3} is intended to derive and analyze the electron-photon dressed states in the presence of electromagnetic field with both linear and circular polarizations. We have derived the electron-light interaction Hamiltonian, found and thoroughly analyzed the energy bandstructure of the irradiated dressed states both analytical and numerically. The final conclusions and outlook are discussed in Section \ref{sec4}.

\section{Model and electronic states in 1T$^\prime$-MoS$_2$}
\label{sec2}

Let us first define our units and estimate the sought quantities related to a real-life experiment. We define our units for the 
energy and momentum as the Fermi energy in graphene for an experimentally accessible yet quite large electron density $n_e^{(0)} = 10^{11}\,
cm^{-2}$.  This corresponds to the Fermi momentum $k_F^{(0)} = \sqrt{2 \pi n^{(0)}} = 7.92 \cdot 10^7 \, m^{-1}$ and the Fermi energy
$E_F^{(0)} = \hbar v_F k_F^{(0)} = 8.32 \cdot 10^{-21}\,J = 52.02 \, meV$. Thus, our unit of length is obtained as $l^{(0)} = 1/k_F^{(0)} = 1.26 \cdot 10^{-8}\, m \backsim 10\,nm$. 

\medskip 
The logical starting point of building our model would be presenting the low-energy Hamiltonian and the electronic states for 1T$^\prime$-MoS$_2$. In the vicinity of two inequivalent $K$ and $K^{\prime}$ points, corresponding to $\xi = \pm 1$ the main Hamiltonian 

\begin{equation}
\label{MainHam}
\hat{\mc{H}}_{\xi=\pm 1}^{\,1T^{\prime}} ({\bf k}) = V_1 k_x \, \Gamma^{(2,0)} + \left\{
- \xi V_- \Gamma^{(0,0)} + \xi V_+ \, \Gamma^{(3,0)} + V_2 \, \Gamma^{(1,1)}  
\right\} \, k_y + \Delta_0 \left[
\xi \Gamma^{(2,0)} - i r_E \, \Gamma^{(2,0)} \cdot \Gamma^{(3,0)}
\right]
\end{equation}
is linear which means the 1T$^\prime$-MoS$_2$ belong to Dirac materials. Here, the components of anisotropic Fermi velocity 
$V_ = 0.286$, $V_+ = 0.721$, $V_1 = 0.387$ and $V_2 = 0.046$ are given in terms of $v_F = 10^{6}\, m/s$ for 
graphene. The spin-orbit coupling gap is $\Delta_0 = 0.81\,E^{(0)}$. $r_E = E_z/E_c$ is the relative value of 
the our-of-plane electric field and $E_c$ is its critical value which closes the bandgap in 1T$^\prime$-MoS$_2$. 

\par 
Following Ref.~[\onlinecite{tan2021anisotropic}], Hamiltonian \eqref{MainHam} is written in terms of $4 \times 4$ gamma matrices $\Gamma^{(0,0)} = \tau_0 \otimes \sigma_0$ is a $4 \times 4$ unit matrix, $\Gamma^{(1,1)} = \tau_1 \otimes \sigma_1$, 
$\Gamma^{(2,0)} = \tau_2 \otimes \sigma_0$ and $\Gamma^{(3,0)} = \tau_3 \otimes \sigma_0$, where $\otimes$ means outer 
product (or Kronecker product) and $\tau_i$ and $\sigma_i$ are regular $2 \times 2$ Pauli matrices acting in pseudospin and 
real-spin spaces, correspondingly.   

\begin{figure} 
\centering
\includegraphics[width=0.95\textwidth]{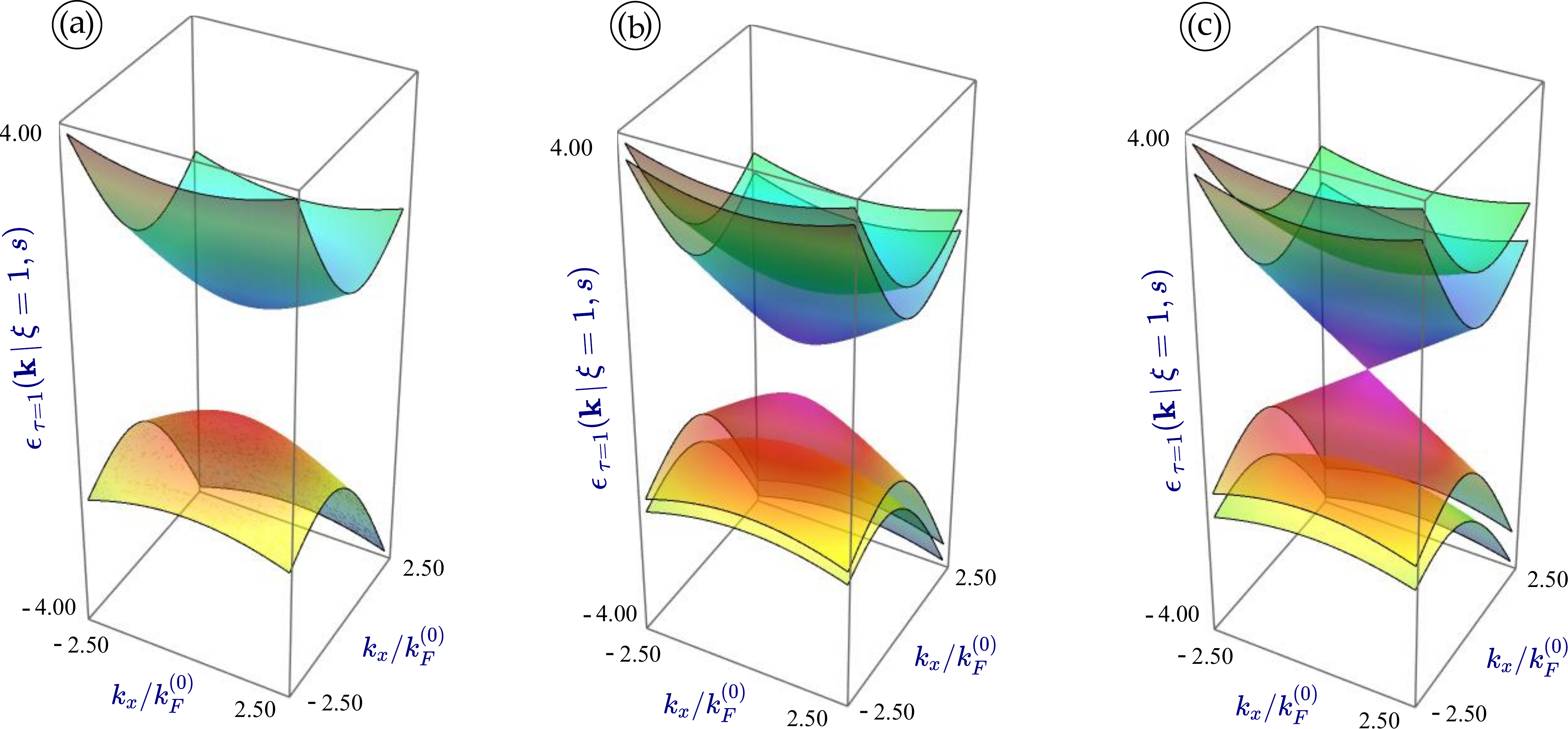}
\caption{(Color online)  Valley- and spin-polarized energy dispersions $\epsilon_{\tau=\pm 1}({\bf k} \, \vert \, \xi=1, s)$ for 1T$^\prime$-MoS$_2$ with a finite spin-orbit coupling gap $\Delta_0$ in the absence of dressing field. Panels $(a)$, $(b)$ and $(c)$ correspond to different values of 
the external electric field $r_E = 0.0$, $0.5$ and $1.0$, as labeled. We have chosen valley index $\xi = 1$ for all the presented cases. 
}
\label{FIG:1}
\end{figure}

In the matrix form, we can rewrite Hamiltonian \eqref{MainHam} as

\begin{equation}
\hat{\mc{H}}_{\xi=\pm 1}^{\,1T^{\prime}} ({\bf k}) = \left\{
\begin{array}{cccc}
- \xi \, (V_+ + V_-) k_y & 0 & r_E \Delta_0 - i V_1 k_x & \xi \Delta + V_2 k_y \\
0 & - \xi \, (V_+ + V_-) k_y &  \xi \Delta + V_2 k_y & r_E \Delta_0 - i V_1 k_x \\
0 & 0 &  \xi \, (V_+ - V_-) k_y & 0 \\
0 & 0 & 0 &  \xi \, (V_+ - V_-) k_y 
\end{array}
\right\} + h.c. \, ,
\end{equation} 
where $h.c.$ means a Hermitian conjugate matrix. The low-energy dispersions in the vicinity of two Dirac points are immediately 
obtained as 

\begin{eqnarray}
\label{disp0}
&& \epsilon_{\tau=\pm 1}({\bf k} \, \vert \, \xi, s) = - \xi \, V_- k_y + \tau \,\mbb{S}({\bf k} \, \vert \, \xi, s) \, , \\
&& \mbb{S}({\bf k} \, \vert \, \xi, s) = \sqrt{\left[ (\xi - s \, r_E) \Delta_0 + V_2 k_y \right]^2 +(V_+ k_y)^2 + (V_1 k_x)^2} \, , 
\end{eqnarray}
where $\tau=\pm 1$ described the electron/hole states related to the conduction and valence bands and $s=\pm 1$ is the real spin index. We immediately discern that the obtained energy spectrum is anisotropic and tilted relatively to the $k_y$ axis ( $- \xi \, V_- k_y$ terms gives the contributions equal and opposite sing around $k_y = 0$).  The energy bandgap is indirect and closes if $\xi - r_E$. Similarly to silicene,  $\xi > s \, r_E$ is a topological insulator phase and $\xi < s \, r_E$ is a regular band insulator. 

\medskip 

The corresponding wave functions are

\begin{equation}
\Psi_{\xi=\pm 1}^{\,1T^{\prime}} ({\bf k})  = \left\{
\begin{array}{c}
s \, \mc{D}_{\,s,\,\xi}({\bf k}, \Delta_0)   \\
- \mc{D}_{\,s,\,\xi}({\bf k}, \Delta_0)   \\
1 \\
-s
\end{array}
\right\} \, \tet{e}^{i \xi k_x x} \, \tet{e}^{i k_y y} \, , 
\end{equation}
where 

\begin{equation}
\mc{D}_{\,s,\,\xi}({\bf k}, \Delta_0) = \left\{  
(\xi - s r_E) \, \Delta_0 + V_2 k_y - i s V_1 k_x 
\right\}^{-1} \, \left[ \xi V_+ k_y - \tau \, \mbb{S}({\bf k} \, \vert \, \xi, s)
\right] \,
\end{equation}
and $\mbb{S}({\bf k} \, \vert \, \xi, s)$ was defined in Eq.~\eqref{disp0}.

\begin{figure} 
\centering
\includegraphics[width=0.65\textwidth]{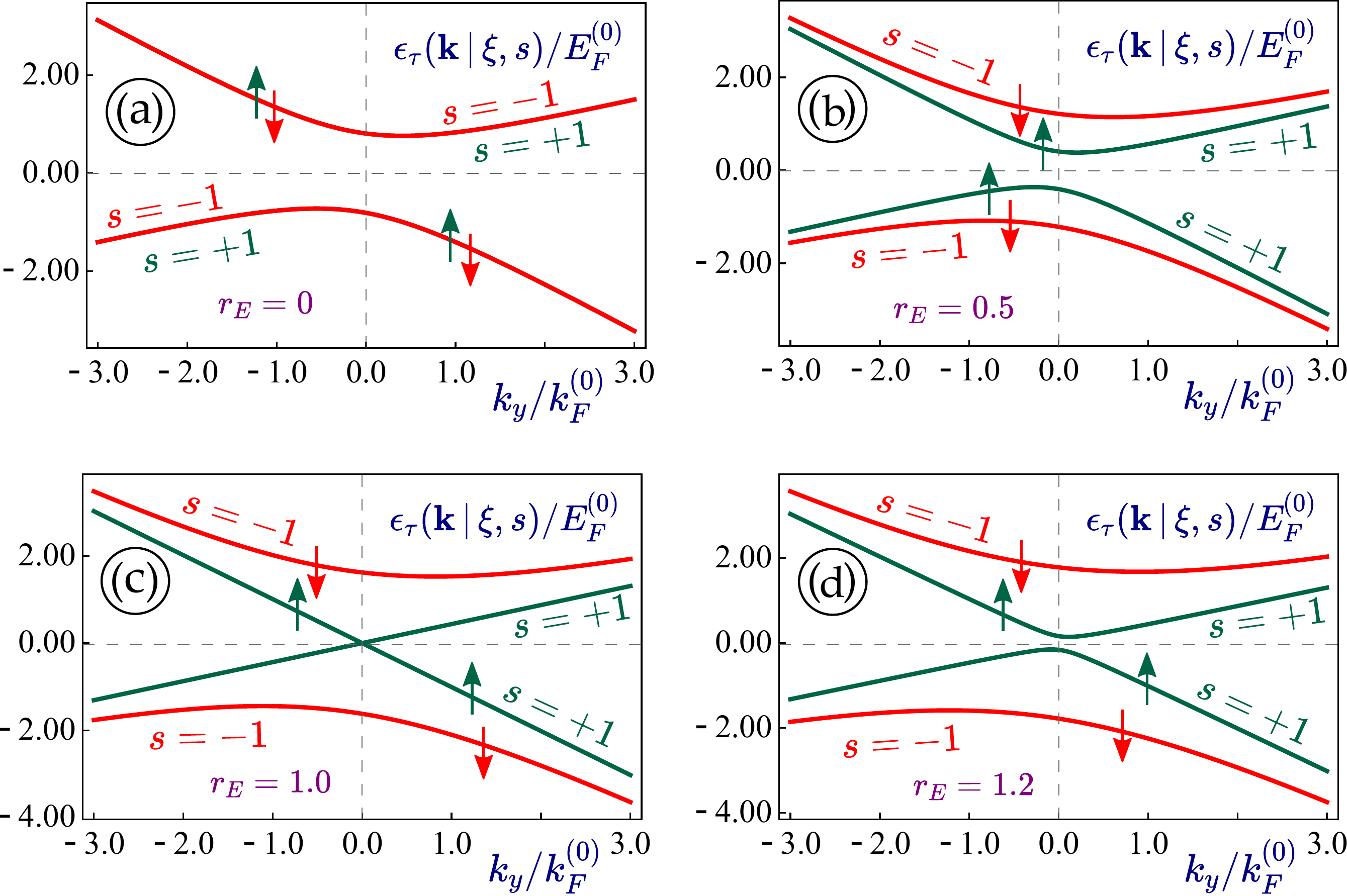}
\caption{(Color online) $k_y$-dependence of the energy dispersions $\epsilon_{\tau=\pm 1}(k_x=0, k_y \, \vert \, \xi, s)$ for non-irradiated 
1T$^\prime$-MoS$_2$. Each panel corresponds to a different values of the perpendicular electric field $r_E = 0.0$, $0.5$, $1.0$ (for which the smaller bandgap for spin $s=1$ is closed) and $r_E = 1.2$, according to our labels. We take $\xi=1$ and $k_x=0$ for all considered cases. 
}
\label{FIG:2}
\end{figure}

\begin{figure} 
\centering
\includegraphics[width=0.65\textwidth]{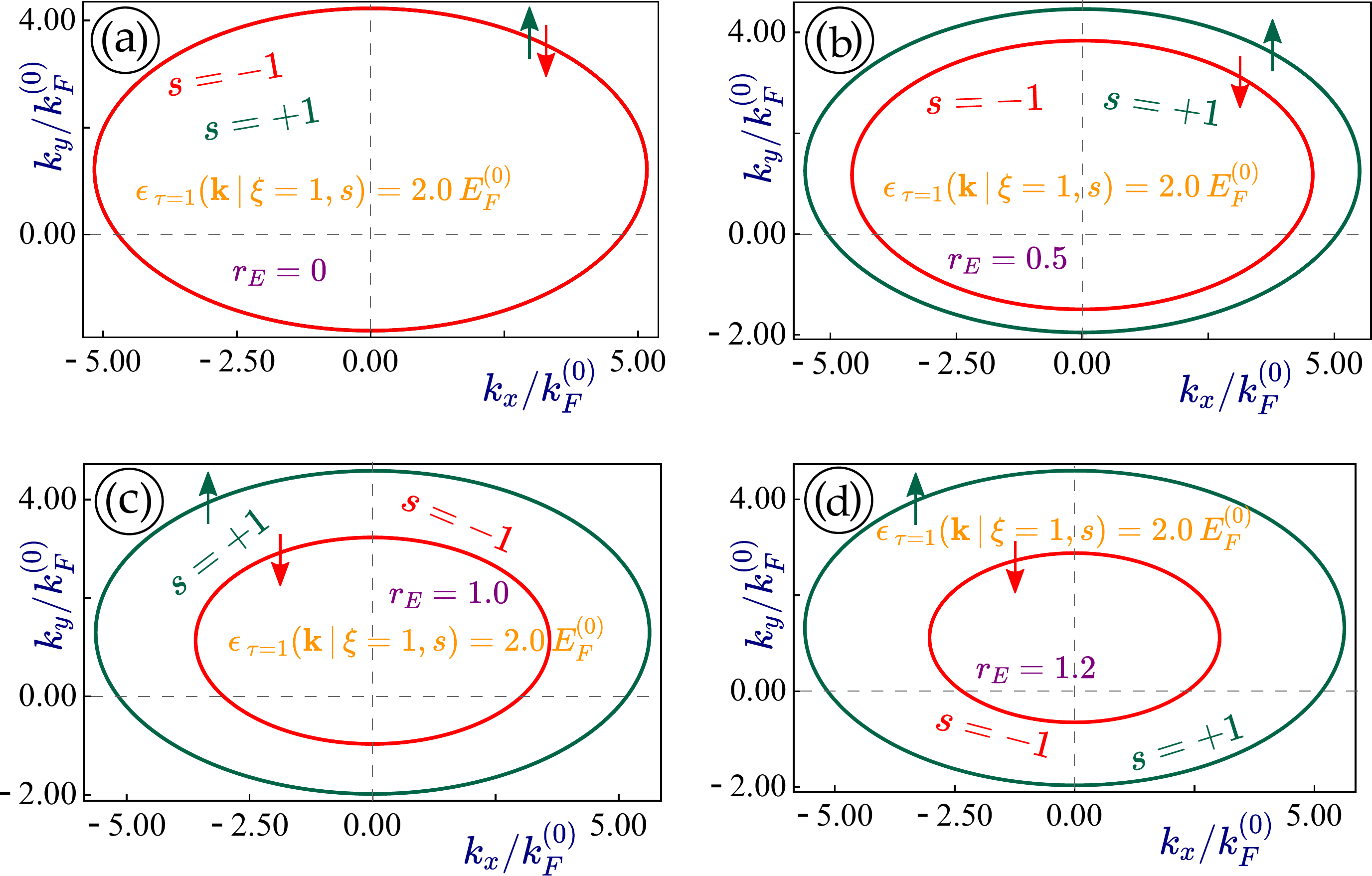}
\caption{(Color online) Angular dependence of the constant-energy $\varepsilon_0 = 2.0\,E^{(0)}$ cut of the 
dispersions $\epsilon_{\tau=\pm 1}(k_x=0, k_y \, \vert \, \xi, s) = 2.0\,E^{(0)}$ for non-irradiated 1T$^\prime$-MoS$_2$. 
Each panel corresponds to a different values of the perpendicular electric field $r_E = 0.0$, $0.5$, $1.0$ and $1.2$.
We have chosen $\xi = 1$ for all plots. 
}
\label{FIG:3}
\end{figure}

First, we should say that we encounter a highly unusual electronic states because even though the Hamiltonian is linear to the components of the wave vector $k_x$ and $k_y$, the Fermi velocities $V_+$ and $V_-$ are present along the main diagonal of the Hamiltonian and the gap terms which stay finite for ${\bf k} = 0$ are located off-diagonal as well. This Hamiltonian structure looks highly unusual and was not encountered in any of the previously considered materials such as graphene, silicene, regular transition metal dichalcogenides, 8-pmmn borophene or even phosphorenes. 

The obtained energy dispersions, presented in Figs.~\ref{FIG:1} and \ref{FIG:2} are tilted and anisotropic. There is no mirror symmetry for $k_y \leftrightarrow -k_y$ while such symmetry is obviously present for $k_x$ component of the electron momentum, which is very well seen from Fig.~\ref{FIG:4}. In general, we have four non- equivalent Fermi velocities $V_+$, $V_-$, $V_1$ and $V_2$ which affect different properties of 1T$^\prime$-MoS$_2$ and contribute to its anisotropy.  

The two subbands in each of the valence and conduction bands depend on the spin index $s=\pm 1$. Thus, we observe two inequivalent subbands and two different bandgaps unless there is no external perpendicular electrostatic field and $r_E = 0$.  The smaller bandgap could be closed for $r_E = \xi = \pm 1$ and our lattice becomes semi metallic. For $\vert r_E \vert <  1$, the band structure represents a topological insulator, and for $\vert r_E \vert > 1$ - regular band insulator. This situation is similar to silicene. All the dispersions and band gaps depend on the valley Index $\xi = \pm 1$ which makes our electronic states spin- and- valley- polarized.

\section{Electron dressed states}
\label{sec3}

In this Section, we aim to calculating and analyze the new electronic states originating as a result of 
the electron-photon interaction due to an applied high-frequency external optical field. These modified 
electronic states are often defined as electron dressed states and the field is referred to as dressing 
field. 

\medskip

The vector potential of the dressing field enters effective Hamiltonian through the canonical substitution
of the wave vector  ${\bf k}$ which enter Hamiltonian \eqref{MainHam} $k_{i} \longrightarrow  k_{i} - e A_i^{(P)}(t)$,
where the time-dependent vector potential of applied field $A_i^{(P)}$ is mainly determined by its polarization.

The driven-induced gauge fields substantially renormalize the band gaps and the spin-orbit splitting of
our investigated material. In order to solve the obtained eigenvalue problem, we must rely on a
perturbative Floquet-Magnus expansion of the interaction Hamiltonian in powers of inverse
frequency which works particularly well for an off-resonant high-frequency irradiation. \,\cite{goldman2014periodically}

Most general type is elliptical polarization (clockwise): 

\begin{equation}
\label{ellipA}
\mbox{\boldmath$A$}^{(E)}(t) = 
\left[  \begin{array}{c}
          A^{(E)}_x (t) \\
          A^{(E)}_y (t)
        \end{array}
\right] = \frac{E_0}{\omega} \left\{
\begin{array}{c}
\cos \Theta^{(p)} \cos (\omega t) - \beta \, \sin \Theta^{(p)} \sin (\omega t) \\ 
\sin \Theta^{(p)} \cos (\omega t) + \beta \,\cos \Theta^{(p)} \cos (\omega t)
\end{array}
\right\} \, ,
\end{equation}
where polarization angle is $\Theta^{(p)}$. The vector potential for a linearly polarized light corresponds to 
$\beta = 0$ in \eqref{ellipA}

\begin{equation}
\label{linA}
\mbox{\boldmath$A$}^{(L)}(t) = 
\left[  \begin{array}{c}
          A^{(L)}_x (t) \\
          A^{(L)}_y (t)
        \end{array}
\right] = \frac{E_0}{\omega} \left\{
\begin{array}{c}
\cos \Theta^{(p)} \\
\sin \Theta^{(p)}
\end{array}
\right\} \, \cos (\omega t ) \ .
\end{equation}

Since the Hamiltonian in Eq.\eqref{MainHam} is linear in $k_{x,y}$, in the presence of $\mbox{\boldmath$A$}^{(L)}(t)$ it only acquires an additional {\it interaction} term, yielding

\begin{equation}
\label{Tlinham}
\hat{\mc{H}}_{1} (\mbox{\boldmath$k$}\, \vert \, \tau) \Longrightarrow \hat{\mbb{H}}^{(L)}(\mbox{\boldmath$k$}, t) = 
\hat{\mc{H}}_{1} (\mbox{\boldmath$k$}\, \vert \, \tau) + \hat{\mc{H}}_A^{(L)}(t) \ , 
\end{equation}
where the $\mbox{\boldmath$k$}$ independent interaction term is

\begin{eqnarray}
 \label{HAL}
 && \hat{\mc{H}}_A^{(L)}(t) = c_0 \, \cos (\omega t)  \times \\
 \nonumber 
 && \times \left\{
 \begin{array}{cccc}
  \xi \left[ V_+ + c_- \right] \sin \Theta^{(p)} & 0 & i V_1 \cos \Theta^{(p)} & - V_2 \sin \Theta^{(p)} \\
  0 &  \xi \left[ V_+ + V_- \right] \sin \Theta^{(p)} & - V_2 \sin \Theta^{(p)} & i V_1 \cos \Theta^{(p)} \\
  0 & 0 &  \xi \left[ V_- - V_+ \right] \sin \Theta^{(p)} & 0 \\  
	0 & 0 & 0 &  \xi \left[ V_- - V_+ \right] \sin \Theta^{(p)} 
 \end{array}
 \right\} + h.c. \ 
\end{eqnarray}
Here we notice an important difference between tilted 1T$^\prime$-MoS$_2$ and all the previously studied materials - the optical-coupling constants $c_0^{(i)} = e V_i E_0/\omega$, corresponding to the different Fermi velocities, are not the same for various elements of Hamiltonian \eqref{HAL}. The only time-dependent term found in matrix \eqref{HAL} is $\backsimeq \cos (\omega t)$ and is related to the linearly polarized type of the irradiation, not to a specific material.

\begin{figure} 
\centering
\includegraphics[width=0.65\textwidth]{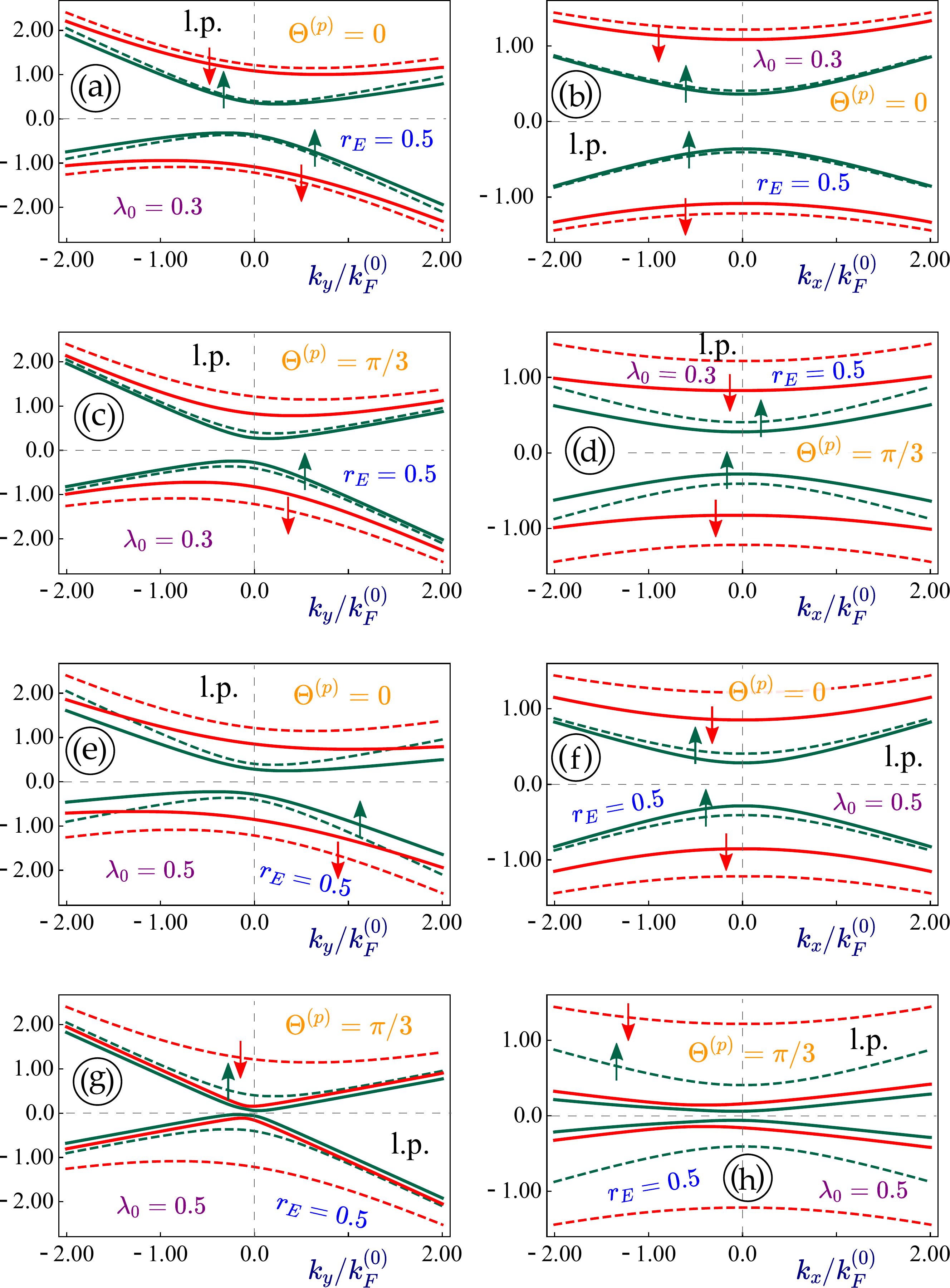}
\caption{(Color online) Energy dispersions $\epsilon^{(L)}_{\tau=\pm 1}(\lambda_0, {\bf k} \, \vert \, \xi, s)$ for the case of linearly polarized irradiation applied to a 1T$^\prime$-MoS$_2$ lattice. The corresponding dispersions for non-irradiated materials and the same values of the lattice parameters are shown as dashed lines for comparison. Each of the panels corresponds to a specific value of the electron-light coupling parameter $\lambda_0 = 0.3$ and $0.5$ and the direction of the linear polarization of the dressing field relative to the $x-$ axis $\Theta^{(P)}$ in accordance with our plot labels. All the left panels $(a)$, $(c)$, $(e)$ and $(g)$ describe the energy dispersions as a function of the $k_y$ component of the wavevector, while the right ones $(b)$, $(d)$, $(f)$ and $(g)$ show their $k_x$-dependence. The vertical electric field $r_E = 0.5$ (when both energy bandgaps are open) is selected for all plots. 
}
\label{FIG:4}
\end{figure}

\begin{figure} 
\centering
\includegraphics[width=0.65\textwidth]{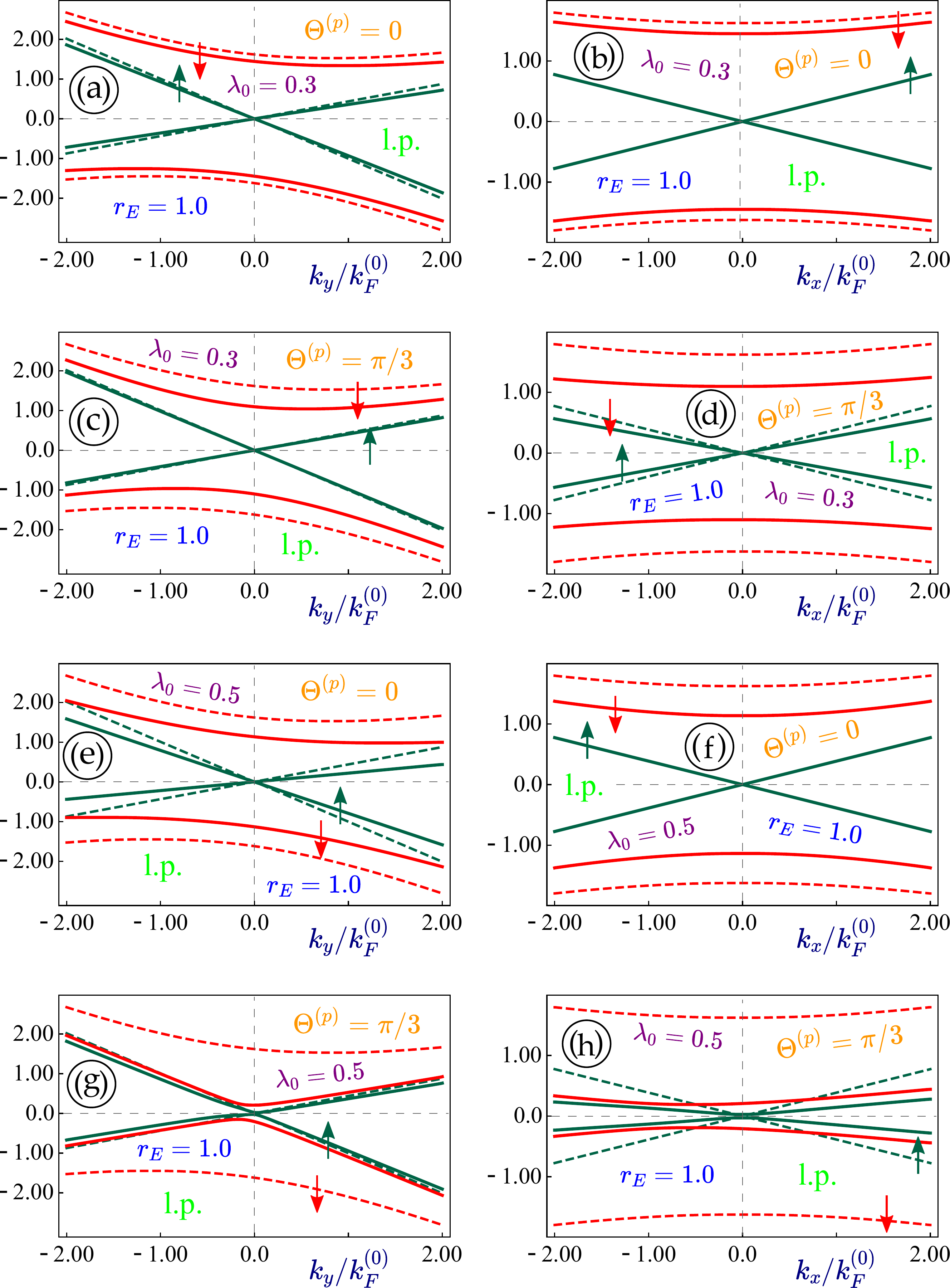}
\caption{(Color online) Energy dispersions $\epsilon^{(L)}_{\tau=\pm 1}(\lambda_0, {\bf k} \, \vert \, \xi, s)$ for the case of linearly polarized irradiation applied to a 1T$^\prime$-MoS$_2$ lattice. The corresponding dispersions for non-irradiated materials and the same values of the lattice parameters are shown as dashed lines for comparison. Each of the panels corresponds to a specific value of the electron-light coupling parameter $\lambda_0 = 0.3$ and $0.5$ and the direction of the linear polarization of the dressing field relative to the $x-$ axis $\Theta^{(P)}$ in accordance with our plot labels. All the left panels $(a)$, $(c)$, $(e)$ and $(g)$ describe the energy dispersions as a function of the $k_y$ component of the wavevector, while the right ones $(b)$, $(d)$, $(f)$ and $(g)$ show their $k_x$-dependence. The vertical electric field $r_E = 1.0$ (when the smaller energy bandgap is closed) is selected for all plots.}
\label{FIG:5}
\end{figure}

\par 
If the direction of the linear polarization is along the $x-$axis, $\Theta^{(p)} = 0$ and Hamiltonian \eqref{HAL} is reduced to

\begin{equation}
 \label{HALx}
\hat{\mc{H}}_A^{(L)}(t) = i \, c_0 V_1  \, \cos (\omega t)   
\, \left[
 \begin{array}{cccc}
  0 & 0 & 1 & 0 \\
  0 &  0 & 0 & 1 \\
  -1 & 0 & 0 & 0 \\  
	0 & -1 & 0 & 0
 \end{array}
 \right]  = - \frac{c_0 \, V_1}{2}  \cos (\omega t) \, \Gamma^{(2,0)} \, ,    
\end{equation}
where we use the following notations: $c_0 = v_F e E_0/\omega$ ($E_0$ is the amplitude of the electric field of our dressing irradiation) and $\lambda_0 = c_0/(\hbar \omega)$.

\medskip

The Floquet-Magnus perturbation approach which has been widely used to obtain the electronic states in the presence of a 
off-resonance and high-frequency dressing field \,\cite{goldman2014periodically} requires that the only time-dependent second term in Eq.\,\eqref{HAL} is rewritten as

\begin{equation}
\hat{\mc{H}}_A^{(L)}(t)  = \hat{\mbb{O}}_1(c_0, \tau) \, \tet{e}^{i \omega t} + 
\hat{\mbb{O}}_1^{\dagger}(c_0, \tau) \, \tet{e}^{-i \omega t} \ , 
\end{equation}
in which matrix $\hat{\mbb{O}}_1(c_0, \tau)$ is free from time dependence

 \begin{eqnarray}
\label{OL}
&& \hat{\mbb{O}}_1(c_0, \tau) =  \frac{c_0}{2} \times \\
 \nonumber 
&& \times \left\{
 \begin{array}{cccc}
  \xi \left[ V_+ + V_- \right] \sin \Theta^{(p)} & 0 & i V_1 \cos \Theta^{(p)} & - V_2 \sin \Theta^{(p)} \\
  0 &  \xi \left[ V_+ + V_- \right] \sin \Theta^{(p)} & - V_2 \sin \Theta^{(p)} & i V_1 \cos \Theta^{(p)} \\
  0 & 0 &  \xi \left[ V_- - V_+ \right] \sin \Theta^{(p)} & 0 \\  
	0 & 0 & 0 &  \xi \left[ V_- - V_+ \right] \sin \Theta^{(p)} 
 \end{array}
 \right\} + h.c. = \\
\nonumber 
&& = \xi \, c_0 \, \sin \Theta^{(p)} \left[ V_+ \Gamma^{(3,0)} -V_- \Gamma^{(0,0)} \right] -  c_0  V_1 \cos \Theta^{(p)} \,
\Gamma^{(2,0)} - c_0 V_2 \sin \Theta^{(p)} \, \Gamma^{(1,1)} \, . 
\end{eqnarray}

Using the Floquet-Magnus series expansion over the powers of $1/(\hbar \omega)$, the effective Hamiltonian for the dressed state 
is obtained as

\begin{eqnarray}
\nonumber
\hat{\mc{H}}_{\text{eff}}^{\,(L)}(\mbox{\boldmath$k$}\, \vert \, \tau)& =&  \hat{\mc{H}}_{1}(\mbox{\boldmath$k$}\, \vert \, \tau) + \frac{1}{\hbar \omega} \, \left[ \, \hat{\mbb{O}}_1(c_0, \tau) , \, \hat{\mbb{O}}_1^{\dagger}(c_0, \tau)  \,\right]\\
\label{Tmexp}
& + & \frac{1}{2 (\hbar \omega)^2} \left\{
 \left[ \left[
 \, \hat{\mbb{O}}_1(c_0, \tau), \, \hat{\mc{H}}_{\, 1} (\mbox{\boldmath$k$}\, \vert \, \tau) \, 
 \right], \, 
 \hat{\mbb{O}}_1^{\dagger}(c_0, \tau) \,  
 \right]
 \,\, + \,\, h.c.
 \right\} \,\, + \cdots \ .
\end{eqnarray}
As it should be the case with the perturbation expansions, the effective Hamiltonian \eqref{Tmexp} is presented as 
the initial non-perturbed Hamiltonian \eqref{MainHam} and a small correction due to the dressing field. 

\par
The following term $\left[ \, \hat{\mbb{O}}_1(c_0, \tau), \, \hat{\mbb{O}}_1^{\dagger}(c_0, \tau) \right]$ is $k$-independent
and, therefore, largely defines the irradiation-induced bandgap of our dressed states for $k=0$. However, for 1T$^\prime$ MoS$_2$
the consequent expansion terms can also affect the dispersions at ${\bf k} = 0$. 
In the case of linearly polarized light, it is obviously zero since $\hat{\mbb{O}}_1(c_0, \tau)$ in Eq.~\eqref{OL} is Hermitian.

The remaining term in Eq.\,\eqref{Tmexp} is written as $\hat{\mbb{T}}_2(\lambda_0 \, \vert \, k, \theta_{\bf k})$.For $ \Theta^{(p)} = 0$, the 2nd order correction to interaction Hamiltonian is reduced to 

\begin{eqnarray}
\label{t200}
&& \hat{\mbb{T}}_2(\lambda_0 \, \vert \, k, \theta_{\bf k}) = \frac{\lambda_1^2}{4} \, 
\left\{
\begin{array}{cccc}
\xi \, V_+ k_y & 0 & - r_E \Delta_0 & - (\xi \Delta_0 + V_- k_y) \\
0 & 4 \xi \, V_+ k_y & - (\xi \Delta_0 + V_- k_y) & - r_E \Delta_0 \\
0 & 0 & -  \xi \, V_+ k_y & 0 \\
0 & 0 & 0 & -   \xi \, V_+ k_y 
\end{array}
\right\} + h.c. = \\
\nonumber
&& = \frac{\lambda_1^2}{4} \,  \left\{
\xi \, V_+ k_y \, \Gamma^{(3,0)} -  (\xi \Delta_0 + V_- k_y) \, \Gamma^{(1,1)} + r_E \Delta_0 \,  \Gamma^{(2,0)}  
\right\} \ ,
\end{eqnarray}

The energy dispersions under linearly polarized dressing field could be obtained analytically but the expressions are
way too long and complicated, even for $\Theta^{(p)}=0$. Therefore, we will focus on the numerical investigation of the 
obtained dressed states. 

\par 
In the simplest case of $r_E = \xi$ (we also choose $\xi = 1$ for the sake of clarity), both non-irradiate and irradiated energy dispersions have no bandgap and the latter is expressed as $(\lambda_0 =  v_F e E_0/(\hbar \omega^2) \ll 1)$

\begin{equation}
\label{alin1}
\epsilon^{(L)}_{\tau=\pm 1}(\lambda_0, {\bf k} \, \vert \, \xi = 1, s) = - \xi \, V_- k_y \pm \sqrt{
(V_1 k_x)^2 + \left[ 
1 - 2 \lambda_0^2 V_1^2 
\right]^2 \, \left( 
V_1^2 + V_+^2
\right) \, k_y^2 \, .
}
\end{equation} 

Please keep in mind that each of the Fermi velocities $V_1$, $V_2$, $V_+$ and $V_-$ is dimensionless and is given as a ration to the Fermi 
velocity in graphene $v_F = 10^6 \, m/s$.

\medskip

In comparison, a linearly polarized dressing field with the same coupling constant $\lambda_0$ applied to graphene and a dice lattice
leads to dispersions $\pm \hbar v_F \,\sqrt{k_x^2 + a(\lambda_0)^2 k_y^2}$, where $a(\lambda_0) = 1 - \lambda_0^2/2$ for graphene and 
$a(\lambda_0) = 1 - \lambda_0^2/4$ for a dice lattice, correspondingly. 

\begin{figure} 
\centering
\includegraphics[width=0.65\textwidth]{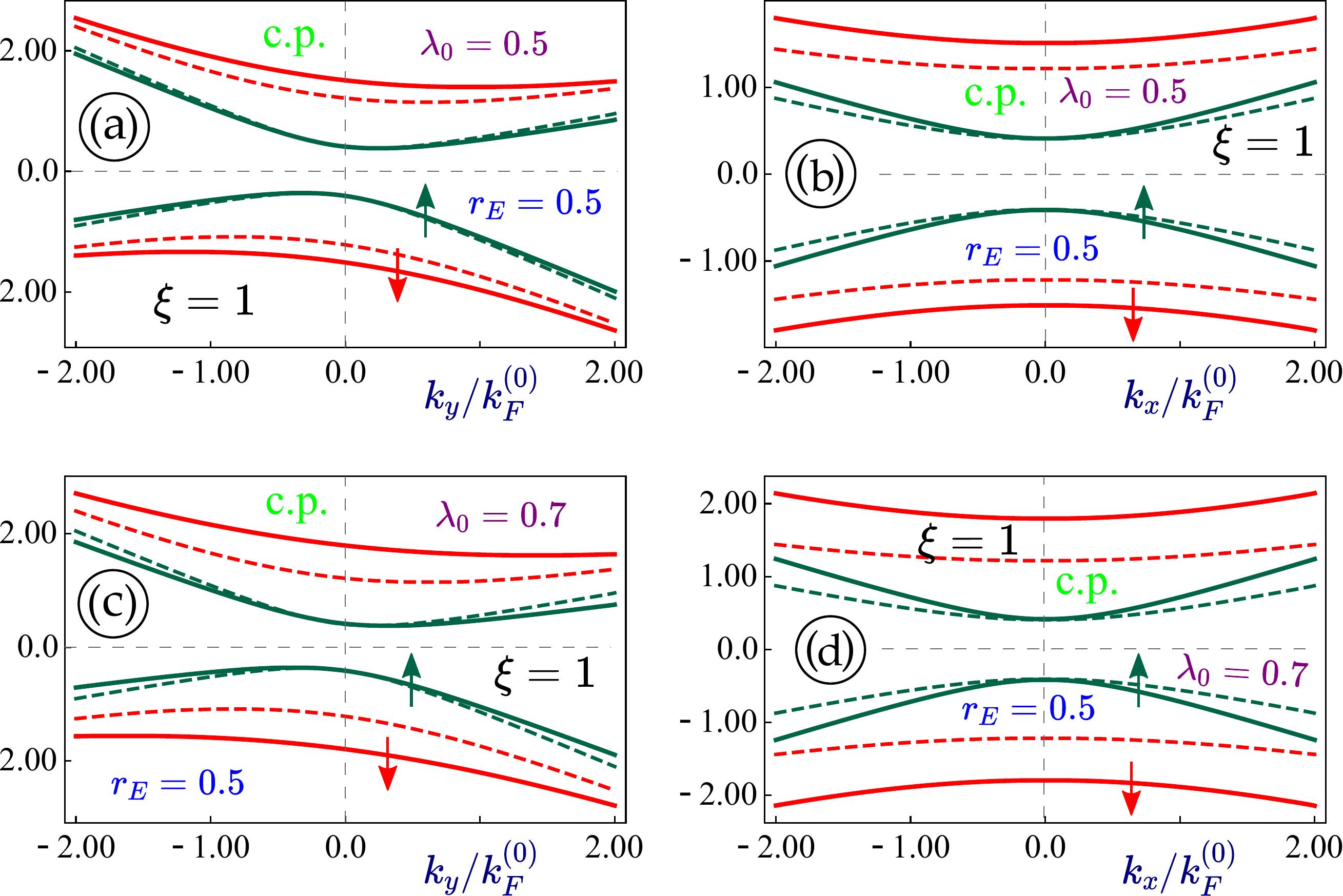}
\caption{(Color online) Energy dispersions $\epsilon^{(C)}_{\tau=\pm 1}(\lambda_0, {\bf k} \, \vert \, \xi, s)$ for the case of circularly polarized dressing field applied to a 1T$^\prime$-MoS$_2$ lattice. The corresponding dispersions for non-irradiated materials and the same values of the lattice parameters are shown as dashed lines for comparison. Each of the panels corresponds to a specific value of the electron-light coupling parameter $\lambda_0 = 0.5$ and $0.7$, as labeled. The left panels $(a)$ and $(c)$, describe the energy dispersions as a function of the $k_y$ component of the wavevector, while the right ones $(b)$ and $(d)$ show their $k_x$-dependence. The vertical electric field $r_E = 0.5$ (both bandgaps are open) and $\xi = 1$ are selected for all plots.}
\label{FIG:6}
\end{figure}

\begin{figure} 
\centering
\includegraphics[width=0.65\textwidth]{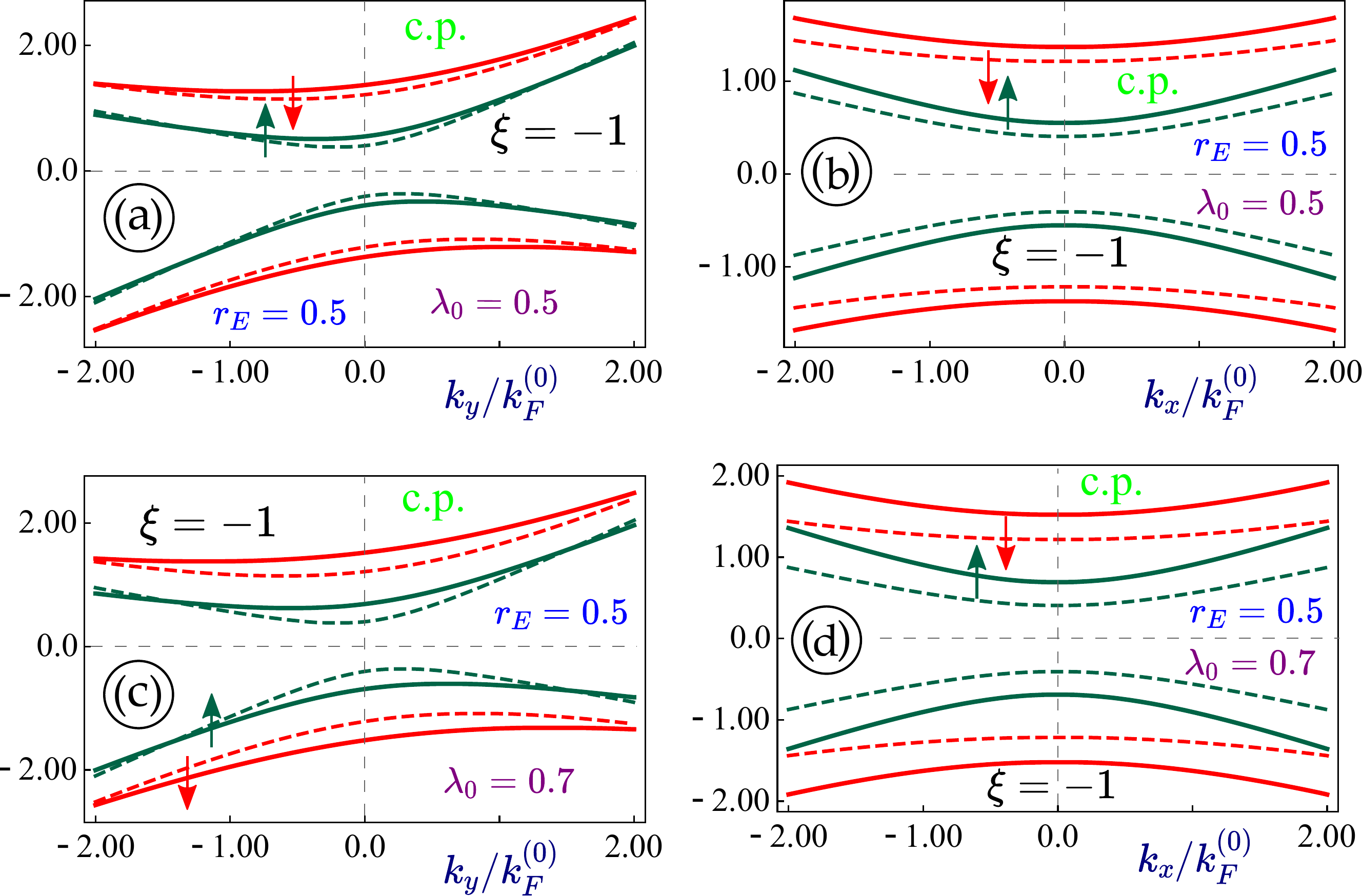}
\caption{(Color online) Energy dispersions $\epsilon^{(C)}_{\tau=\pm 1}(\lambda_0, {\bf k} \, \vert \, \xi, s)$ for the case of circularly polarized dressing field applied to a 1T$^\prime$-MoS$_2$ lattice. The corresponding dispersions for non-irradiated materials and the same values of the lattice parameters are shown as dashed lines for comparison. Each of the panels corresponds to a specific value of the electron-light coupling parameter $\lambda_0 = 0.5$ and $0.7$, as labeled. The left panels $(a)$ and $(c)$, describe the energy dispersions as a function of the $k_y$ component of the wavevector, while the right ones $(b)$ and $(d)$ show their $k_x$-dependence. The vertical electric field $r_E = 0.5$ (both bandgaps are open) and $\xi = - 1$ are selected for all plots.}
\label{FIG:7}
\end{figure}

\begin{figure} 
\centering
\includegraphics[width=0.65\textwidth]{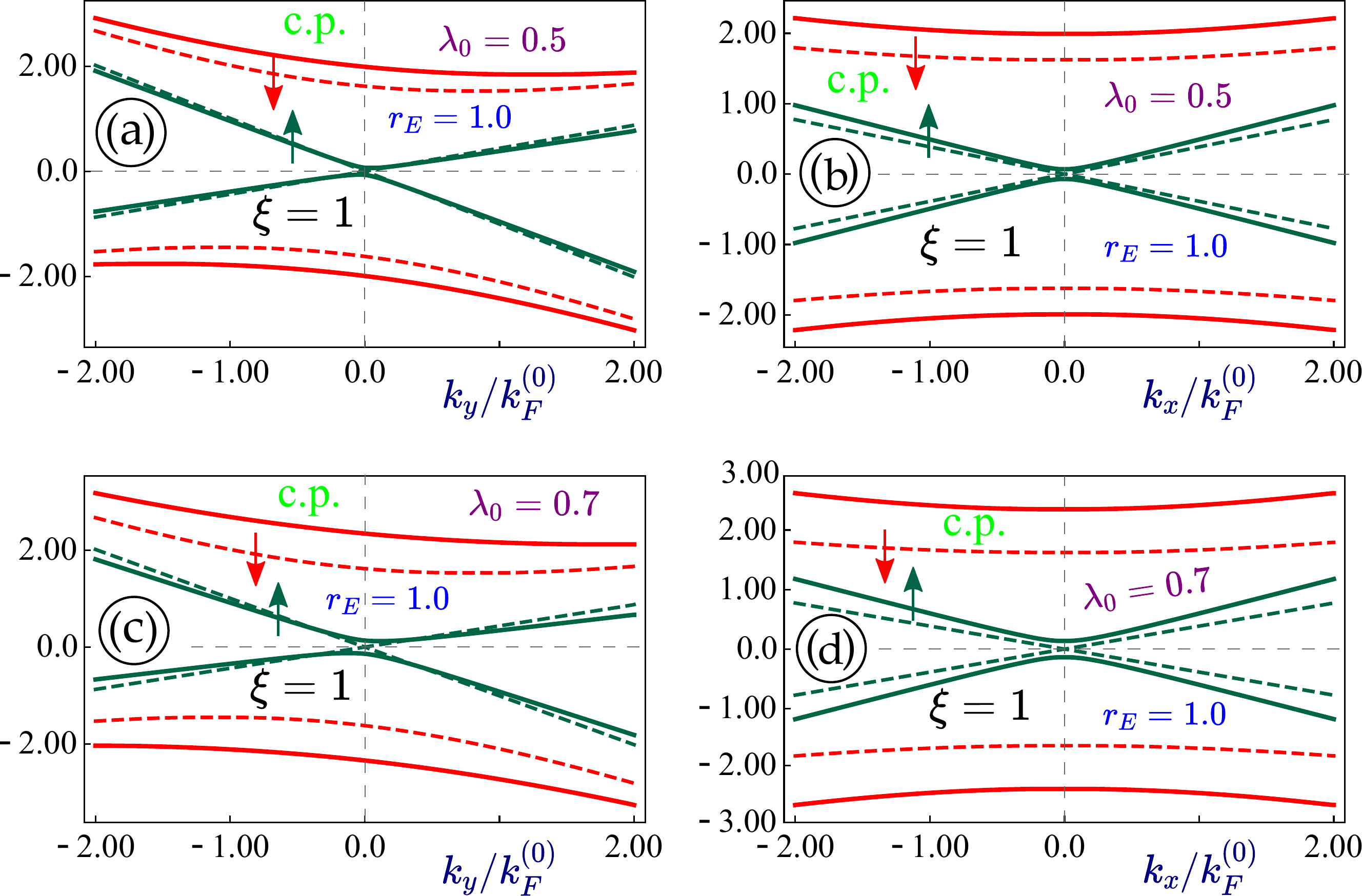}
\caption{(Color online) Energy dispersions $\epsilon^{(C)}_{\tau=\pm 1}(\lambda_0, {\bf k} \, \vert \, \xi, s)$ for the case of circularly polarized dressing field applied to a 1T$^\prime$-MoS$_2$ lattice. The corresponding dispersions for non-irradiated materials and the same values of the lattice parameters are shown as dashed lines for comparison. Each of the panels corresponds to a specific value of the electron-light coupling parameter $\lambda_0 = 0.5$ and $0.7$, as labeled. The left panels $(a)$ and $(c)$, describe the energy dispersions as a function of the $k_y$ component of the wavevector, while the right ones $(b)$ and $(d)$ show their $k_x$-dependence. The vertical electric field $r_E = 1.0$ (semi-metallic dispersions with no bandgap) and $\xi = 1$ are selected for all plots.}
\label{FIG:8}
\end{figure}

\begin{figure} 
\centering
\includegraphics[width=0.65\textwidth]{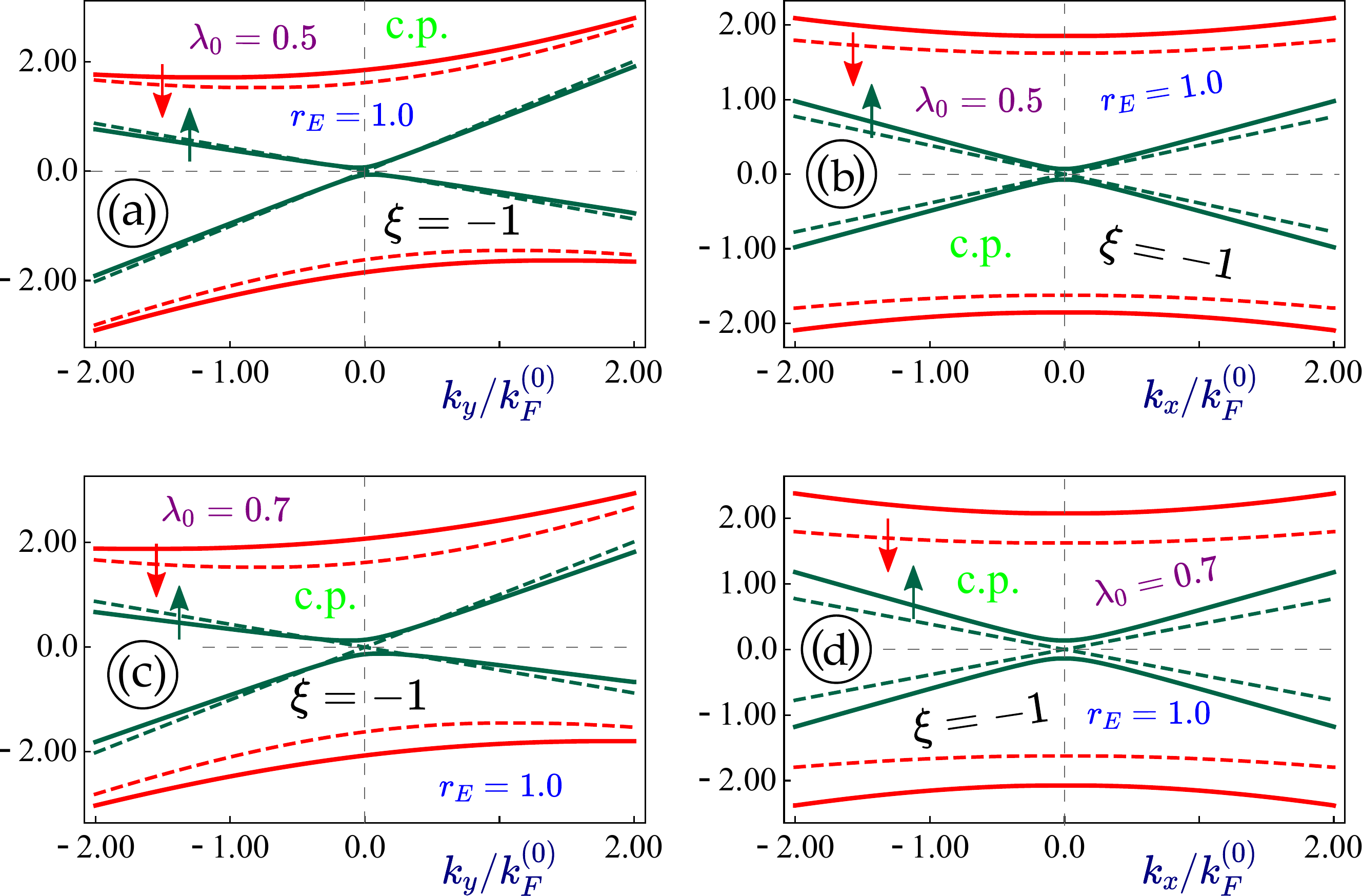}
\caption{(Color online) Energy dispersions $\epsilon^{(C)}_{\tau=\pm 1}(\lambda_0, {\bf k} \, \vert \, \xi, s)$ for the case of circularly polarized dressing field applied to a 1T$^\prime$-MoS$_2$ lattice. The corresponding dispersions for non-irradiated materials and the same values of the lattice parameters are shown as dashed lines for comparison. Each of the panels corresponds to a specific value of the electron-light coupling parameter $\lambda_0 = 0.5$ and $0.7$, as labeled. The left panels $(a)$ and $(c)$, describe the energy dispersions as a function of the $k_y$ component of the wavevector, while the right ones $(b)$ and $(d)$ show their $k_x$-dependence. The vertical electric field $r_E = 1.0$ (semi-metallic dispersions with no bandgap) and $\xi = 1$ are selected for all plots.}
\label{FIG:9}
\end{figure}

 Linearly polarized irradiation is generally known to induce or modify the existing anisotropy in two-dimensional lattices. If the initial bandgap is zero, it stays zero in graphene or an $\alpha-\mc{T}_3$. Since we are dealing with initially anisotropic lattice, the direction of the light polarization matters, and the obtained energy dispersions are essentially different for the different directions $\Theta^{(p)}$ of the linear polarization of the dressing field. 

We consider the states with  $r_E = 0.5$ and a finite bandgap and present our results in Fig.~\ref{FIG:4}, as well as initially gapless bandstructure shown in Fig.~\ref{FIG:5}. Here, we compare how the energy dispersions are modified in the presence of a dressing field (solid lines) compared to the non-irradiated states (dashed lines).  

We clearly see that all the important characteristics of the energy dispersions are affected by the linearly polarized light. Unlike graphene, both the direct and indirect energy bandgaps are decreased when the off-resonant dressing field is applied.  

Tilting, anisotropy and $k_x \leftrightarrow - k_x$ mirror symmetry of the electronic states are affected are affected by the linearly polarized irradiation. The locations of the critical points (minima and maxima) of each subband are shifted as well. We also note that energy subbands corresponding to the different values of spin index values are affected differently by linearly polarized light. 

\par
For $\Theta^{(p)}$ the dispersions in both $x-$ and $y-$ in-plane directions are modified and we see a lot richer physical picture which includes breaking the $k_x  \leftrightarrow - k_x$  mirror symmetry of the dressed states. We also see that for the semi-metallic states with 
$r_E = \xi$ for which one of the bandgaps is closed, it also remains zero under linearly polarized light. The slopes of the dispersions, however, are affected differently for the $x-$ and $y-$ directions revealing field-induced anisotropy, just as we observed for graphene. The 
$k_x \leftrightarrow -k_x$ mirror symmetry remains unaffected by such field only for $\Theta^{(p)} = 0$.  

\subsection{Circularly polarized light}

The opposite limit $\beta = 1$ is related to circular polarization

\begin{equation}
\label{circA1}
\mbox{\boldmath$A$}^{(C)}(t) = 
\left[  \begin{array}{c}
          A^{(C)}_x (t) \\
          A^{(C)}_y (t)
        \end{array}
\right] = \frac{E_0}{\omega} \left\{
\begin{array}{c}
\cos \left[ \Theta^{(p)} + \omega t \right] \\
\sin \left[ \Theta^{(p)}  + \omega t \right]
\end{array}
\right\} \ .
\end{equation}
Since we are looking for stationary and time-independent states, the initial phase $\Theta^{(p)}$ for circularly polarized field 
could be removed with no loss of generality.
\par 
Thus, the vector potential for circularly polarized irradiation is simplified as 

\begin{equation}
\label{circA}
\mbox{\boldmath$A$}^{(C)}(t) = 
\left[  \begin{array}{c}
          A^{(C)}_x (t) \\
          A^{(C)}_y (t)
        \end{array}
\right] = \frac{E_0}{\omega} \left\{
\begin{array}{c}
\cos \left( \omega t \right) \\
\sin \left( \omega t \right)
\end{array}
\right\} \ .
\end{equation}

The interaction Hamiltonian for circularly polarized light with vector potential \eqref{circA} becomes

\begin{eqnarray}
 \label{HAC}
 && \hat{\mc{H}}_A^{(C)}(t) =  \\
 \nonumber 
 && = c_0 \, \left\{
 \begin{array}{cccc}
  \xi \left[ V_+ + V_- \right] \sin (\omega t) & 0 & i V_1 \cos (\omega t) & - V_2 \sin (\omega t) \\
  0 &  \xi \left[ V_+ + V_- \right] \sin (\omega t) & - V_2 \sin (\omega t) & i V_1 \cos (\omega t) \\
  0 & 0 &  \xi \left[ V_- - V_+ \right] \sin (\omega t) & 0 \\  
	0 & 0 & 0 &  \xi \left[ V_- - V_+ \right] \sin (\omega t) 
 \end{array}
 \right\} + h.c.  = \\
\nonumber 
&& = c_0 \left\{ \xi  \, \sin (\omega t) \left[ V_- \, \Gamma^{(0,0)} + V_+ \, \Gamma^{(3,0)} \right] - 
V_2 \,  \sin (\omega t) \, \Gamma^{(1,1)} - V_1 \,  \cos (\omega t) \, \Gamma^{(2,0)}
\right\} \, . 
\end{eqnarray}
Using Eq.~\eqref{HAC}, we can immediately obtain the time-independent interaction matrix $\hat{\mbb{O}}_1(c_0, \tau)$ as

\begin{eqnarray}
&& \hat{\mbb{O}}_1(c_0, \xi) = c_0 \left\{ 
- i \xi \, \left[ V_- \, \Gamma^{(0,0)} + V_+ \, \Gamma^{(3,0)} \right] + 
i V_2 \, \Gamma^{(1,1)} -  V_1 \, \Gamma^{(2,0)} 
\right\} = \\
\nonumber 
 && = \frac{- i\, c_0}{2} \, \left\{
 \begin{array}{cccc}
 \xi  \, \left[ V_+ + V_- \right]  & 0 &  - V_1  &  - V_2 \\
 0 &  \xi  \, \left[ V_+ + V_- \right] & - V_2 &  - V_1  \\
 V_1 & - V_2 & \xi \, \left[ V_- - V_+ \right]  & 0 \\  
	- V_2 &  V_1 & 0 & \xi  \, \left[ V_- - V_+  \right]
 \end{array}
 \right\} \, ,
\end{eqnarray}
Now we can easily obtain the 1st order correction $\frac{1}{\hbar \omega} \, \left[ \, \hat{\mbb{O}}_1(c_0, \tau) , \, \hat{\mbb{O}}_1^{\dagger}(c_0, \tau)  \,\right]$ of our series expansion

\begin{equation}
\label{fo}
\frac{1}{\hbar \omega} \, \left[ \, \hat{\mbb{O}}_1(c_0, \tau) , \, \hat{\mbb{O}}_1^{\dagger}(c_0, \tau)  \,\right] = 
\frac{c_0^2 \, V_1}{4 \, \hbar \omega} \, \left\{
 \begin{array}{cccc}
0 &  V_2 & \xi \, V_+ & 0 \\
V_2  & 0 & 0  & \xi \, V_+ \\
\xi \, V_+ &  0 & 0 &  -V_2 \\  
0 &\xi \, V_+  & - V_2 & 0
 \end{array}
\right\} \, . 
\end{equation}

The resulting energy dispersions are obtained as 

\begin{eqnarray}
\label{ecirc}
&& \epsilon_{\tau=\pm 1}({\bf k} \, \vert \, \xi, s) = - \xi \, V_- k_y + \tau \times \\
\nonumber 
&& \times \sqrt{\left[ \, \Delta_{\xi, s}(\lambda_0)^2 + V_2 k_y \right]^2 +(V_+ k_y)^2 + (V_1 k_x)^2} \, , 
\end{eqnarray}

For $k_x=k_y=0$, the energy dispersions reveal spin- and valley-dependent bandgaps

\begin{equation}
\Delta_{\xi, s}(\lambda_0) = \pm \left\{
[(\xi + s \, r_E) \Delta_0]^2 + 2 \lambda_0^2 V_1 V_+ 
\Delta_0 (1 + s \, \xi \, r_E) + \lambda_0^4 V_1^2 (V_2^2 + V_+^2) \,
\right\}^{1/2} \, .
\end{equation}
The finite-$k$ terms of dispersions \eqref{ecirc} are not affected by the circularly polarized irradiation since only first-term correction
term was taken into account to obtain this simplified analytical expressions. 

Finally, we consider the modification of the energy dispersions in the presence of circularly polarized light taking into account the first- and the second-order correction terms and present our numerical results in Figs.~\ref{FIG:6} - \ref{FIG:9}.  

The most interesting observation is definitely that we see a tremendous difference between the two different valleys $\xi = 1$ and $\xi=-1$, shown separately in each of the figures. For $\xi = 1$, the smaller bandgaps remain nearly unaffected, while the larger ones demonstrate a significant increase. Surprisingly, for $\xi = -1$ both subbands and both bandgaps show almost the same increase.

The circularly polarized irradiation normally leads to opening a bandgap in most known Dirac materials. If non-irradiated 1T$^\prime$-MoS$_2$ lattice has zero gap, it becomes finite when a dressing field with circular polarization is applied.

\par

Not only the states with no dressing field, but the irradiation-induced energy gaps demonstrate a strong dependence on the spin and valley indices and, therefore, are spin- and valley-polarized. Another interesting feature of the dressed states under circular polarized radiation is that the initially bigger bandgaps are affected stronger and become even larger in the presence of a dressing field. This property shows a stark difference with graphene and silicene in which the gapless states are mostly modified by circularly polarized irradiation. In 1T$^\prime$-MoS$_2$, the slopes or Fermi velocities are also strongly affected by this type of dressing field.

\section{Summary and remarks}
\label{sec4}

Our present work represents an attempt to construct a new type of low-energy dispersions and electronic states in recently 
discovered distorted tetragonal molybdenum disulfide 1T$^\prime$-MoS$_2$ by applying Floquet engineering.

\par 
Floquet theory and, broadly speaking, Floquet engineering represent one of the most promising and currently studied directions
in the present-day material science and quantum optics. A fundamental interest to this topic which connects various aspects of 
condensed-matter physics stems from the fact that one can foresee some crucial and unusual electronic properties of a novel 
material before it is actually tested in experiment.

\par
The resulting electron dressed state represents a single quasiparticle which combines the crucial features of both irradiated 
matter and the applied dressing field. The properties of these states are mainly determined by the polarization of the applied 
irradiation.

\medskip

We have performed a rigorous theoretical and numerical investigation into the electron dressed states in 
1T$^\prime$-MoS$_2$ in the presence of external radiation with various polarizations, which includes the derivation of 
closed-form analytical expressions for the electron energy band structure driven by external irradiation in the 
terahertz regime. 

\par

We have found that there is no clear distinction between the effects of linearly and circularly polarized light for 1T$^\prime$-MoS$_2$, the obtained physical picture is a lot richer and more involved than what we have seen for any previously considered Dirac materials. Specifically, linearly polarized light which has been mostly noticed to create or modify the in-plane anisotropy of a two-dimensional lattice, in our case also substantially increases the larger bandgap which has not been previously observed.  

\par 

One of our most important findings is that not only the initial states in 1T$^\prime$-MoS$_2$ but also their modification by the dressing field substantially depends on the valley index $\xi = \pm 1$. This includes changing both direct and indirect bandgaps, tilting and Fermi velocities in different directions. Therefore, the obtained dressed states also show significant degree of valley polarization. 

\par

The energy dispersion have been also obtained analytically for both linear and circular polarizations, but in general their expressions are so complicated that we had to choose only very specific cases - fixed valley index $\xi = 1$ and zero band gap for linearly polarized light and only first-order approximation for circular polarized irradiation - to provide our analytical results. However, these approximated equations give us a pretty good understanding of how the electron states are modified by the irradiation. The linearly polarized dressing field modified some of the Fermi velocities and, therefore, leads to a change of the existing anisotropy and tilting, while applying circularly polarized irradiation causes mostly the renormalization of the existing bandgaps.

\medskip 

 Our results for the modified electron dispersions could be later tested
by transmission electron microscopy, scanning tunneling spectroscopy, angle-resolved photoemission
spectroscopy (ARPES). This work is expected to become a noticeable contribution to an unprecedented research effort
on Floquet engineering of all innovative two-dimensional Dirac materials.

\bibliography{FloquetTiltedBib}

\end{document}